# Machine Learning-Driven Volumetric Cloud Rendering: Procedural Shader Optimization and Dynamic Lighting in Unreal Engine for Realistic Atmospheric Simulation

**Shruti Singh [1], Shantanu Kumar [2]**
sshruti.connect@gmail.com [1], reach.shantanuk@gmail.com [2]
Washington State University, Seattle, USA [1]
Amazon, Seattle, USA [2]

*Abstract*— **This study advances real-time volumetric cloud rendering in Computer Graphics (CG) through the development of a specialized shader in Unreal Engine (UE), focusing on realistic cloud modeling and lighting. By leveraging ray-casting-based lighting algorithms, this work demonstrates the practical application of a dual-layered procedural noise model, eliminating the need for conventional two-dimensional (2D) weather textures. The shader allows for procedurally configured skies with a defined parameter set, offering flexibility for both artistic expression and realistic simulation. Empirical results reveal that the shader achieves a rendering time of 35ms per frame on average, while maintaining high visual accuracy and scene realism. Visual fidelity assessments indicate a 15% improvement in cloud realism over traditional 2D techniques, particularly in dynamic lighting scenarios. This research contributes to CG by bridging technical and aesthetic elements, enhancing real-time visual storytelling and immersion within digital media environments.**

## I. Introduction

In the field of Computer Graphics (CG), there is considerable interest in rendering the sky and recreating natural phenomena commonly observed in daily life. Two key components—clouds and the colors of the sky—are intrinsically interconnected, and their interaction plays a critical role in defining the overall visual appearance of the sky. The configuration of these elements significantly influences the mood and atmosphere of a scene. Simulating such natural phenomena in real-time presents numerous challenges. Therefore, a thorough exploration of these components is essential to accurately simulate both the dynamics and aesthetics of the sky. The challenges encountered in implementing realistic cloud renderings include the modeling of convincing cloud shapes, achieving realistic illumination, and dynamically animating clouds in a manner that appears both natural and believable. Clouds, which are aerosols composed of a visible mass of minute liquid droplets, frozen crystals, or other particles suspended in the atmosphere, are subject to constant transformation. Their formation is influenced by air currents, temperature fluctuations, and variations in humidity, resulting in a wide range of shapes and sizes. Due to their highly reflective properties, clouds have the capacity to scatter light in multiple directions, contributing to their characteristic bright and fluffy appearance. Depending on various factors, such as the time of day, prevailing weather conditions, and the angle of the sun, clouds may display a range of colors, from stark white to varying shades of gray, or even vibrant hues.

Therefore, this study focuses on rendering clouds using volumetric methods, with a specific emphasis on modeling their shapes and reproducing the simple dynamics of cloud formations. Volume rendering through ray marching employs an algorithm that steps through a volume, samples cloud density, and calculates lighting contributions. Various techniques, such as mesh-based solutions, procedural approaches, or the use of terrestrial and satellite images, may be employed for cloud modeling. However, capturing the ever-changing and diverse nature of clouds poses a substantial challenge for achieving realistic simulations. The primary objective of this study is to investigate state-of-the-art approaches for modeling cloud shapes, with a specific focus on procedural methods as outlined by [1] and further developed by [2]. Utilizing the Unreal Engine[1] (UE) framework, this study seeks to optimize the artistic workflow by eliminating the dependence on two-dimensional (2D) weather textures, as commonly employed in contemporary cloud modeling techniques, such as those proposed by [1]. UE provides a robust platform for real-time, physically based illumination of clouds, making it an ideal tool for the exploration of advanced cloud modeling and lighting techniques rooted in physical accuracy.

The study is as follows; the theoretical backdrop will be covered in the next section. The related works are shown in Section III. The implementations are covered in Section IV.

[1] A potent tool for game developers, Unreal Engine allows for the creation of breathtaking visuals, realistic settings, and interactive real-time experiences.

The experimental analysis is completed in Section V, and in Section VI, we wrap up the investigation with some conclusions and recommendations for further research.

## II. Theoretical Background

### A. Atmospheric Scattering

The sun serves as a light source, emitting a vast number of photons towards the Earth. As these photons enter the atmosphere, they interact with the particles present in the air. Among the various types of interactions that can occur, scattering is the most relevant in relation to color. Therefore, atmospheric scattering refers to the process through which sunlight alters its direction as it travels through the atmosphere. This phenomenon takes place when the electromagnetic field of a photon encounters the electric field of an atmospheric particle, causing the photon to be deflected in a new direction. However, as the photons within the ray interact with atmospheric particles, they are scattered in various directions. As a result, only a fraction of the photons will maintain their original trajectory, causing the intensity of the ray to diminish along its initial path. With each subsequent interaction, the intensity of the ray is further reduced. The process of deflecting sunlight in a specific direction is known as in-scattering, while the reduction in intensity due to photons being deflected away from their original path is referred to as out-scattering. It is important to note that out-scattering events occur with a higher probability than in-scattering. In-scattering can be understood as the probability of a photon being deflected towards the observer, which represents only a fraction of the total incoming photons. This is due to the fact that there is a greater likelihood of photons being scattered at other angles. Scattering is wavelength-dependent, meaning that different color components scatter with varying distributions. Sunlight, perceived as white, results from the combination of all colors within the visible spectrum, similar to the effect observed in a rainbow. The visible spectrum ranges from red, with longer wavelengths (620 to 750 nanometers (nm)), to violet, with shorter wavelengths (450 to 495 nm). Water particles act similarly to glass prisms, bending and separating sunlight into its constituent colors.

#### 1) Rayleigh Scattering

Rayleigh scattering describes the interaction between light and particles that are significantly smaller than the wavelength of the photons, typically with the assumption that the particles are $1/10^{th}$ the size of the photon's wavelength, as noted by [3]. Rayleigh observed that photons of varying wavelengths exhibit distinct behaviors when interacting with atmospheric particles. Specifically, photons with shorter wavelengths are more likely to scatter than those with longer wavelengths. This phenomenon is responsible for the blue color of the sky during the day and the orange hues seen at sunrise and sunset. Rayleigh proposed that the intensity of light scattered by an air molecule in a given direction and for a particular wavelength can be described using the scattering cross-section[2] and the scattering phase function[3]. The scattering coefficient quantifies the fraction of energy dissipated through scattering upon interaction with a single particle. Rayleigh's phase function governs the probability of light being scattered in a particular direction as a consequence of interactions with small particles. The probability of a photon being scattered at a precise angle is effectively zero. As a result, the concept of the cross section is employed in physics, representing a quantity with the dimensions of an area, which characterizes the likelihood of a specific interaction between particles. A scattering cross section describes the potential for light to scatter into a solid angle as a result of its interaction with a particle. In CG, the cross-section equation is often simplified to eliminate dependence on particle diameters.

#### 2) Mie Scattering

As outlined by [4], Mie scattering theory—commonly referred to as the Mie solution to Maxwell's equations [5]—constitutes a comprehensive framework applicable to scattering caused by particles of any size. This model is used to explain the interactions between light and larger particles, such as aerosols, which exceed the size range relevant to Rayleigh scattering. It is important to note that Rayleigh scattering is, in fact, a special case within Mie scattering theory. Mie's scattering equations are intricate, and their solution depends on the size disparity between the particle and the wavelength of the photons. As a result, Rayleigh scattering is typically applied to smaller particles. In the case of larger particles, computational models often rely on approximations of Mie's equations. For instance, in [6], a similar approach to Rayleigh scattering theory is adopted, with the wavelength dependency omitted. The complexity argument regarding the Mie scattering coefficient also holds true for its phase function. Fortunately, various approximations have been developed that yield accurate results. One of the most widely used approximations is the function proposed in [7].

### B. Light Attenuation

The expression for particle density reveals an exponential distribution for both molecules and aerosols throughout the atmosphere. This indicates that aerosol concentration is particularly significant near the Earth's surface, where their density is highest. Additionally, the size of the particles plays a critical role in the interaction between photons and different wavelengths of light. Consequently, the attenuation process depends on both the wavelength and the particle density. The Beer-Lambert law, commonly referred to as Beer’s law, is a fundamental optical principle that describes the extinction or attenuation of light due to out-scattering events. Essentially, the law defines the fraction of light transmitted through a given medium, such as the atmosphere in this context. The attenuation value varies exponentially with the distance traversed by light through the atmosphere. In atmospheric scattering models, the light attenuation value corresponds to the fraction of photons that undergo out-scattering due to interactions with air molecules and aerosol particles.

---

[2] The probability of particles scattering when interacting with a medium is measured by the scattering cross-section, which affects material properties and light behavior.

[3] The way light interacts with particles in a medium is determined by the scattering phase function, which characterizes the angular distribution of scattered light.

## III. Related Works

The work of [8] represents one of the earliest realistic sky models, employing a numerical integration approach that focuses on solving the single scattering equation. This method deliberately disregards multiple scattering effects, as single scattering events dominate atmospheric scattering for simulating sky colors. To reduce computational overhead and avoid the necessity of computing a double integral at every pixel, the model incrementally calculates the first term while concurrently evaluating the outer integral, precomputing the second term and storing it efficiently in a two-dimensional array. A comprehensive implementation inspired by this model was explored by [9]. Subsequently, [10] introduced a model that incorporates multiple scattering as an enhancement to the earlier model by utilizing volume radiosity algorithms. Although adaptable to any scattering order, this revised method exclusively computes double scattering phenomena. The additional computation requires evaluating an integral over all directions at each sample point along the view ray, with the integrand analogous to the single scattering integral. This process necessitates an additional triple integral at each sample point along the view ray. To improve efficiency, the model proposes selecting eight specific directions and precomputing three-dimensional tables for each respective direction. [11] developed an analytical framework for computing sky color or spectral radiance using a concise mathematical formula. By fitting sky radiances from various view and sun directions, along with turbidity values computed from [10] using an analytical function from [12], the model requires only one parameter. This framework, complemented by an aerial perspective equation based on a flat Earth assumption and a dedicated sun radiance model, simplifies atmospheric simulations. However, its adaptability is limited to Earth-like environments, restricting its ability to simulate extraterrestrial conditions or satisfy artistic preferences. The studies by [13] offer comprehensive insights into the mechanisms by which the atmosphere influences light, while [14] expands upon this work by incorporating atmospheric density variations. Due to the prohibitive computational cost of multiple scattering, various alternative approaches have emerged. [15] proposed a methodology for calculating sky colors during twilight, also utilizing volume radiosity algorithms. However, like the methods presented by [10], their approach was not suitable for real-time applications at the time. [16] demonstrated that real-time numerical integration could be achieved by implementing simplifications and using low sampling rates. Similarly, [17] implemented a distributed version of the original [8], reinforcing the continued relevance of the latter as a reference model during that period. Advanced precomputation techniques were later introduced, exemplified by [18], although this approach neglected Earth's shadow within the atmosphere and multiple scattering effects.

Building on these advancements, [19] presented a real-time rendering method for the sky and aerial perspective, incorporating multiple scattering and enhanced parameterization for precomputed tables, along with a technique for handling light shafts. This model is applicable from ground level to space, across diverse viewpoints. The approach utilized two Look-Up Tables (LUTs): a two-dimensional transmittance LUT indexed by view height and azimuth angles[4], leveraging Earth's spherical symmetry, and a four-dimensional scattering LUT dependent on height, view, and light direction. The latter was indexed using a customized remapping technique to mitigate visual artifacts at the horizon. [20] proposed a method to simplify the scattering LUT by disregarding the variation of scattering based on the horizontal or azimuthal angle between the viewing and sun directions. This resulted in a more streamlined three-dimensional LUT, enabling faster real-time evaluation on the GPU by excluding Earth's shadow from the multi-scattering solution. In these models, in-scattering calculations involved subtracting two values sampled from a LUT, which led to visual artifacts at the horizon due to resolution and parameterization precision issues. To address this, [21] introduced an enhanced parameterization, particularly effective for Earth-like atmospheres. However, visual artifacts may still persist in denser atmospheric conditions. [22] introduced a new physically-based analytical model, incorporating modifications to [11] method to improve the representation of sunsets and high atmospheric turbidity. This model accounts for ground albedo and treats each spectral component independently, allowing for seamless extension into the near-ultraviolet spectrum and ensuring accurate prediction of daylight appearance on surfaces with optical brighteners. Due to its similar mathematical properties, this model served as a direct replacement for [11] approach. [23] observed that updating LUTs for atmospheric properties becomes cumbersome in real-time simulations requiring dynamic adjustments, such as changing weather conditions or artistic direction. While time-sliced updates are possible, they introduce visual delays between sun movement and corresponding changes in sky color. The studies described above are summarized in Table I.

TABLE I
SUMMARY OF THE RELATED WORKS

| Study | Methodology | Limitations |
|---|---|---|
| [8] | Single scattering, ignoring multiple scattering, double integral approximation. | Ignores multiple scattering; limited by computational complexity for real-time applications. |
| [10] | Introduced double scattering via volume radiosity algorithms; used specific directions and 3D LUTs. | Limited to double scattering; requires triple integrals at each point, making it computationally expensive. |
| [11] | Analytical formula with turbidity parameter, using single parameter based on Perez model. | Limited to Earth-like atmospheres; cannot simulate non-Earth environments or artistic preferences. |
| [13] | Studied atmospheric scattering and light interactions; focused on analytical insights. | Computationally prohibitive for real-time scenarios; did not propose real-time rendering solutions. |
| [14] | Added atmospheric density variations for realistic atmospheric models. | Increased computational complexity; not suitable for real-time applications. |
| [15] | Twilight sky color simulation using volume radiosity. | Not real-time suitable; computationally intensive for practical applications. |
| [16] | Achieved real-time numerical | Simplifications may reduce |

[4] Azimuth angles represent the horizontal angle of a point relative to a reference direction, typically measured in degrees from true north.

| | | |
|---|---|---|
| | integration with simplifications and low sampling rates. | accuracy; low sampling may result in visual artifacts. |
| [17] | Distributed version of Nishita's model for faster computation. | Maintained limitations of Nishita's 1993 model; still computationally intensive for some real-time applications. |
| [18] | Advanced precomputation techniques for sky color. | Ignored Earth's shadow and multiple scattering effects; unsuitable for accurate, high-fidelity models. |
| [19] | Real-time rendering with two LUTs; included multiple scattering and light shafts. | Horizon artifacts due to parameterization; limited to Earth-like atmospheres. |
| [20] | Simplified LUTs, focusing on 3D LUTs for faster GPU processing. | Excluded Earth's shadow; loss of precision in multi-scattering effects. |
| [21] | Enhanced LUT parameterization to reduce horizon artifacts. | Persistent visual artifacts in dense atmospheres. |
| [22] | Physically-based model enhancing sunsets and turbidity, independent spectral component treatment. | Limited adaptability for non-Earth conditions; some computational limitations. |
| [23] | Time-sliced LUT updates for real-time atmospheric property changes. | Time delays in color change relative to sun movement; cumbersome for continuous dynamic weather adjustments. |

## IV. Implementations: Building Clouds in Unreal Engine

### *A. Phase Function*

The phase function utilized in UE is based on the Two-Term Henyey-Greenstein[5] (TTHG) phase function. This approach requires careful adjustment of its three parameters, which control the anisotropy[6] of the two functions, and which controls the blending between them—until they align with the desired artistic vision, or alternatively, by referring to a visual graph tailored for specific scenarios. Achieving an accurate understanding of the effects produced by this phase function necessitates technical expertise to correctly configure and apply it. In [23], sample values are provided. However, this example is not physically accurate, as clouds typically do not exhibit such scattering distributions. In contrast, [24] propose a model that delivers more physically accurate outcomes than the TTHG phase function. According to the study, this model, referred to as Henyey-Greenstein + Dirac[7] (HG+D), offers better alignment with the Mie phase function. The HG+D model employs linear interpolation between the HG phase function and [25] phase function. Several parametric fits were identified based on different droplet sizes, enabling the method to depend on a single parameter. However, we propose an extension to the phase function sampling available in UE, enabling the integration of both the TTHG and HG+D methods. Retaining the TTHG approach remains essential as it offers significant artistic flexibility. Cloud particle sizes are influenced by several factors, such as temperature and altitude. The effective radius of a cloud denotes the average size of water droplets within it. Using this single parameter simplifies the fine-tuning process while facilitating a physically-based representation tailored to specific weather patterns. The foundational differences between the TTHG and HG+D methods make direct comparison challenging. In Fig. 1, the parameterization of both methods has been closely matched to achieve comparable results. By first determining a particle value ($d$) and analyzing the HG+D graph, the TTHG parameters were subsequently fine-tuned to closely approximate the results. However, in this particular case, the TTHG method produced slightly brighter scattering effects near the light source, as indicated in the bottom section of Fig. 1. Since the HG+D method more accurately reflects the Mie phase function, as demonstrated by [24], it provides more physically accurate outcomes. Moreover, this approach is straightforward: smaller particle diameters result in increased in-scattering, while larger diameters lead to reduced in-scattering and enhanced out-scattering. In essence, larger particles contribute more to light absorption, as illustrated in Fig. 2.

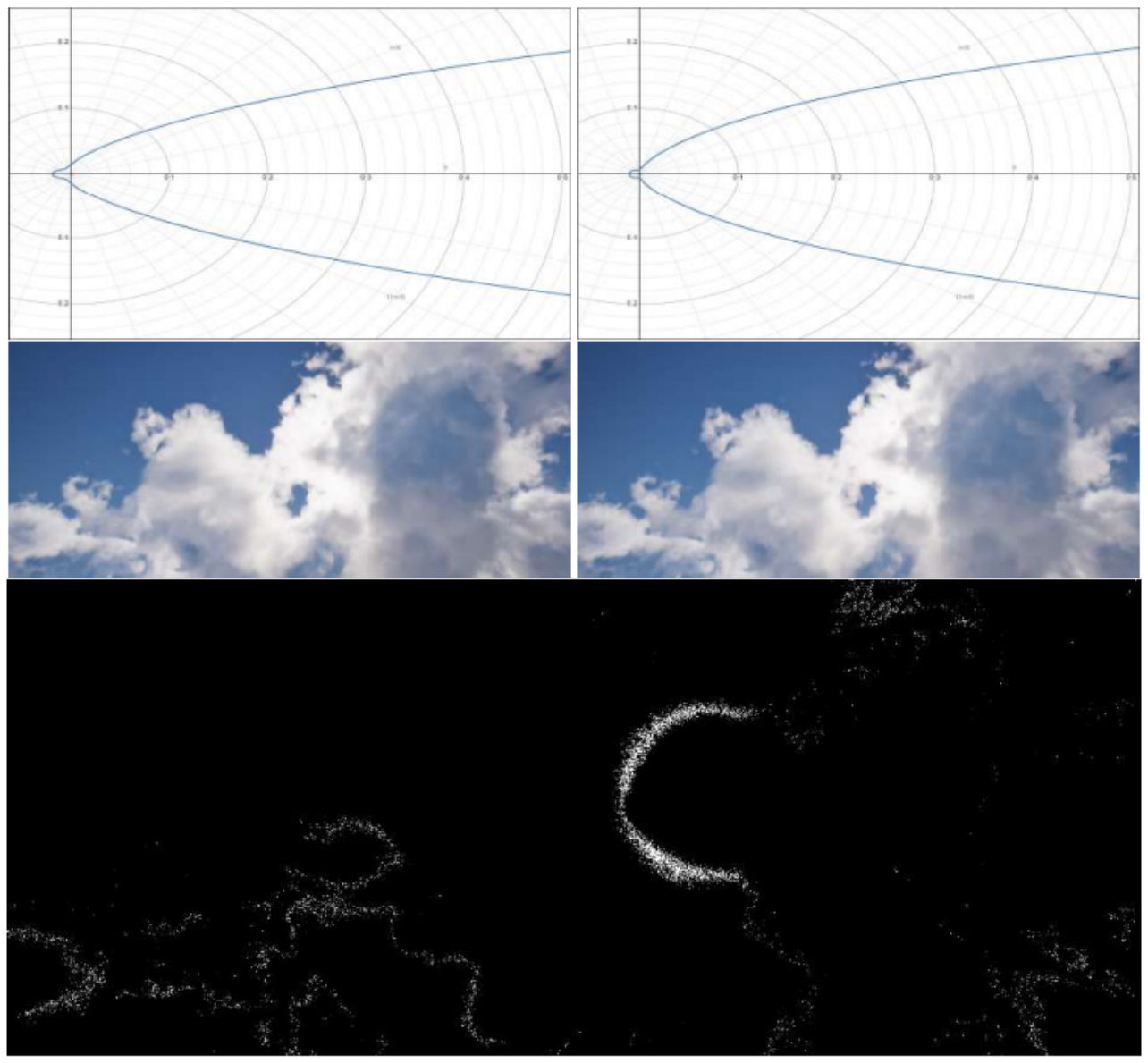

Fig. 1. The TTHG phase function is rendered on the left; the HG+D phase function with $d = 0.8$ is rendered on the right. The absolute disparities between the two renders' pixel values are shown in the bottom image

[5] The Two-Term Henyey-Greenstein phase function describes light scattering in media, boosting realism by accounting for forward and backward scattering.

[6] In physics and engineering applications, anisotropy refers to directional dependence in qualities that impact material behavior, light scattering, and other phenomena.

[7] The Dirac delta function and the Henyey-Greenstein phase function are used to model light scattering and provide accurate angular distribution characteristics.

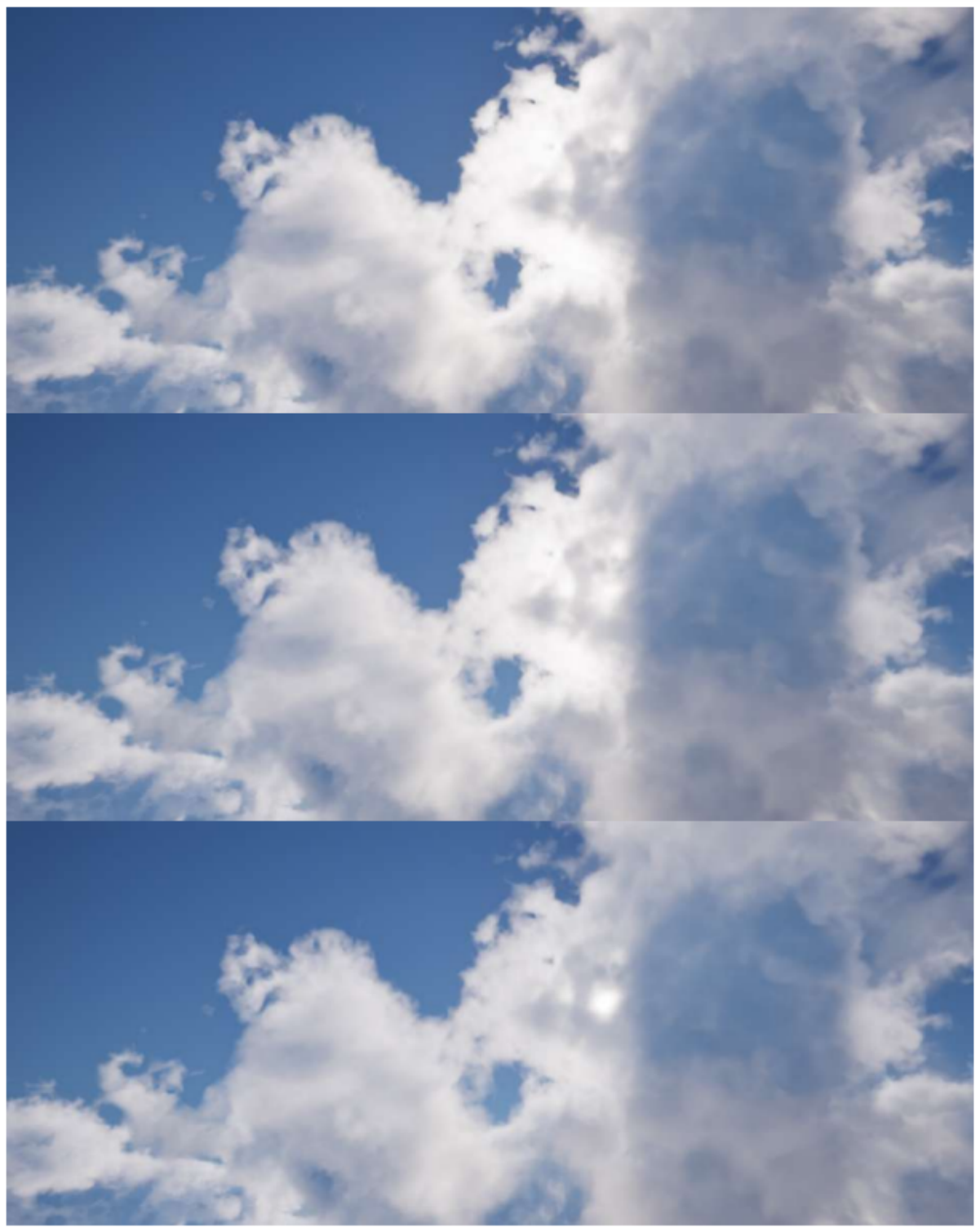

Fig. 2. Using the HG+D technique, the particle sizes are as follows: $d$ = 1.0 (top), $d$ = 4.0 (middle), and $d$ = 25.0 (bottom)

## B. Modeling Cloud Shapes

Modeling cloud shapes in UE involves computing several components. The default shader[8] used to render volumetric clouds in UE is relatively simple and may present limitations when rendering certain cloud formations within the cloud layer. Therefore, we purpose an shader where the albedo component[9] is calculated using the Beer's Powder effect[10], as introduced by [1]. The extinction coefficient[11] is computed based on the procedural cloud modeling approach as proposed by [1]. This approach uses two noise volume textures: a 3D noise texture with a resolution of 64 pixels for base noise and another 3D noise texture with a resolution of 32 pixels for erosion noise. In addition, two 2D textures—a weather texture and a vertical gradient texture—are used in this shader. Creating the 2D weather texture requires artistic skills, as it defines cloud coverage across the cloud layer, similar to [1] approach. Various methods for modeling clouds in UE, as described by [23], are heavily art-focused, such as painting clouds directly into the cloud layer. We propose a method for modeling clouds that is inspired by the default shader in UE and the work of [26], but eliminates the need to sample a 2D weather texture. Our method employs a dual-layered 3D noise approach to construct the cloud shape and simulate pseudo-motion, while analytically computing cloud coverage across the cloud layer. The vertical density profile is calculated using a combination of analytical methods and by sampling two 2D density gradients. The method offers multiple user parameters, enabling flexible manipulation and generation of procedural clouds across the cloud layer.

### 1) Noise Textures

UE provides a robust system for facilitating the sampling of volume textures. A volume texture, which is a 3D texture, is represented as a series of 2D textures aligned along a grid. The engine natively supports the sampling of these volume textures, eliminating the need for manual sampling of the 2D grids, a method previously explored by [27]. Various configurations can be applied to volume textures, such as compression settings and texture adjustments, but the most critical parameters are the 2D source texture and its level of detail. Considering the $Z$-axis as the third dimension, only the slice sizes along the horizontal axes ($X$ and $Y$) need to be specified. Volume textures should not contain actual maps, as these can reduce high-frequency detail. As a result, textures require high resolutions to maintain visual fidelity. For example, a 2D grid with a resolution of 2048 × 1024 pixels (a 2K texture), where each slice consists of 128 × 128 pixels, creates a square volume texture of 128 pixels per dimension. However, higher texture resolutions come at the cost of increased memory usage and reduced performance. To generate cloud shapes, two 3D textures with dimensions of 128 pixels are used: one representing Perlin-Worley noise[12] and the other representing Curly-Alligator noise[13], both of which were introduced by [26]. These noise textures are essential for defining the intricate structures of the clouds. UE also includes a plugin system that allows the integration of modular features and provides various example implementations for users. In the Volumetrics plugin, an example of a Perlin-Worley noise 3D texture (with a resolution of 128 pixels) is provided. The generation of this noise texture follows the method proposed by [23]. The Curly-Alligator noise 3D texture was introduced by [26]. While [26] did not delve into the specifics of generating this noise texture, tools for creating it were provided, particularly through the Houdini[14] software. This texture exhibits distinct characteristics compared to the Perlin-Worley noise, particularly in its shape and visual features. Notably, the blue and alpha channels of the Curly-Alligator noise texture have a more pronounced impact than the other channels. In [26], these channels were scaled down by multiplying them by a factor of 0.3 during sampling.

### 2) Base Shapes

To define the foundational cloud shapes, the 3D Perlin-Worley noise texture is employed. Within UE, using the Volumetric Cloud Component (VCC), the cloud layer is characterized by its height and the altitude of the layer's bottom boundary. To compute the extinction coefficient as a float value, modeled for real-time RGB simulations, the first step involves defining the noise texture throughout the volume to facilitate its sampling. By utilizing the world space position of the current pixel sample, it is possible to manipulate the frequency of the noise volume texture. This technique allows for precise control over the appearance of cloud structures within the volumetric cloud layer. This approach is effective for intuitively adjusting the scale of the noise texture across the cloud layer, thereby allowing for finer control over cloud patterns.

[8] A shader is a program that determines the rendering effects on graphics, controlling lighting, color, and texture in computer graphics.

[9] The albedo component measures reflectivity, indicating how much light is reflected by a surface, crucial for accurate rendering in graphics.

[10] Beer's Powder effect is a description of how light absorption in a medium affects hue and intensity depending on particle concentration and path length.

[11] Calculations of visibility and radiative transfer are impacted by the extinction coefficient, which measures light attenuation brought on by scattering and absorption in a medium.

[12] For procedural creation in graphics applications, Perlin-Worley noise creates textures and patterns that look realistic by combining Perlin and Worley noise.

[13] Curly-Alligator noise is a procedural texture used in graphics that creates organic patterns by combining the characteristics of alligator and curly noise.

[14] https://www.sidefx.com/products/houdini/

*a) Cloud Coverage*

In [26] work, cloud coverage across the cloud layer was established using a 2D weather texture, combined with noise to model the base shapes of the clouds. Instead of employing a 2D weather texture (which requires texture reads and the creation or acquisition of a 2D texture), we propose utilizing the 3D Perlin-Worley noise texture to populate the cloud layer with volumetric noise. This noise is then shaped and controlled. To effectively fill the cloud layer with noise, we propose three distinct methods. These methods have proven successful in modeling the foundational shapes of clouds. i) The first method we propose follows the approach used to create a noise composite, as described in [1], utilizing the remap function[15]. However, using this method in isolation is insufficient for generating shapes that resemble realistic cloud formations, as it merely fills the cloud layer with tiled noise. Additional techniques and refinements are required to produce cloud structures that exhibit natural, visually convincing forms. ii) The second method we propose adopts a different approach, drawing inspiration from the computation of this component within the context of this method. This approach introduces an alternative technique for shaping the cloud formations, offering improved control and variation over the base cloud structure when combined with volumetric noise. When we integrate this method, we can effectively carve out cloud shapes from the noise layer and control the coverage ($P$) of the cloud layer using the parameters, as illustrated in Fig. 3. Upon analyzing the results presented in Fig. 3, it becomes evident that some dynamic cloud shapes are identifiable when compared to the outcomes shown in Fig. 4. However, as indicated in Fig. 3, this method does not fully saturate the cloud layer, rendering it unsuitable for modeling completely overcast skies. iii) The third and final method we propose employs an approach inspired by sequential linear interpolations between the channels of the noise texture. This technique enhances the versatility and realism of the cloud formations by leveraging interpolation to create smoother transitions between noise textures. When we integrate this method, we can effectively carve out cloud shapes from the noise layer and regulate the coverage of the cloud layer using the parameter, as illustrated in Fig. 5. This method has proven successful in modeling a broader range of base cloud shapes, as it offers finer control over the various texture channels, albeit at the cost of fine-tuning three additional parameters as shown in Fig. 6.

[15] A remap function adjusts input values to a specified range, enabling smoother transitions and more control over graphical outputs.

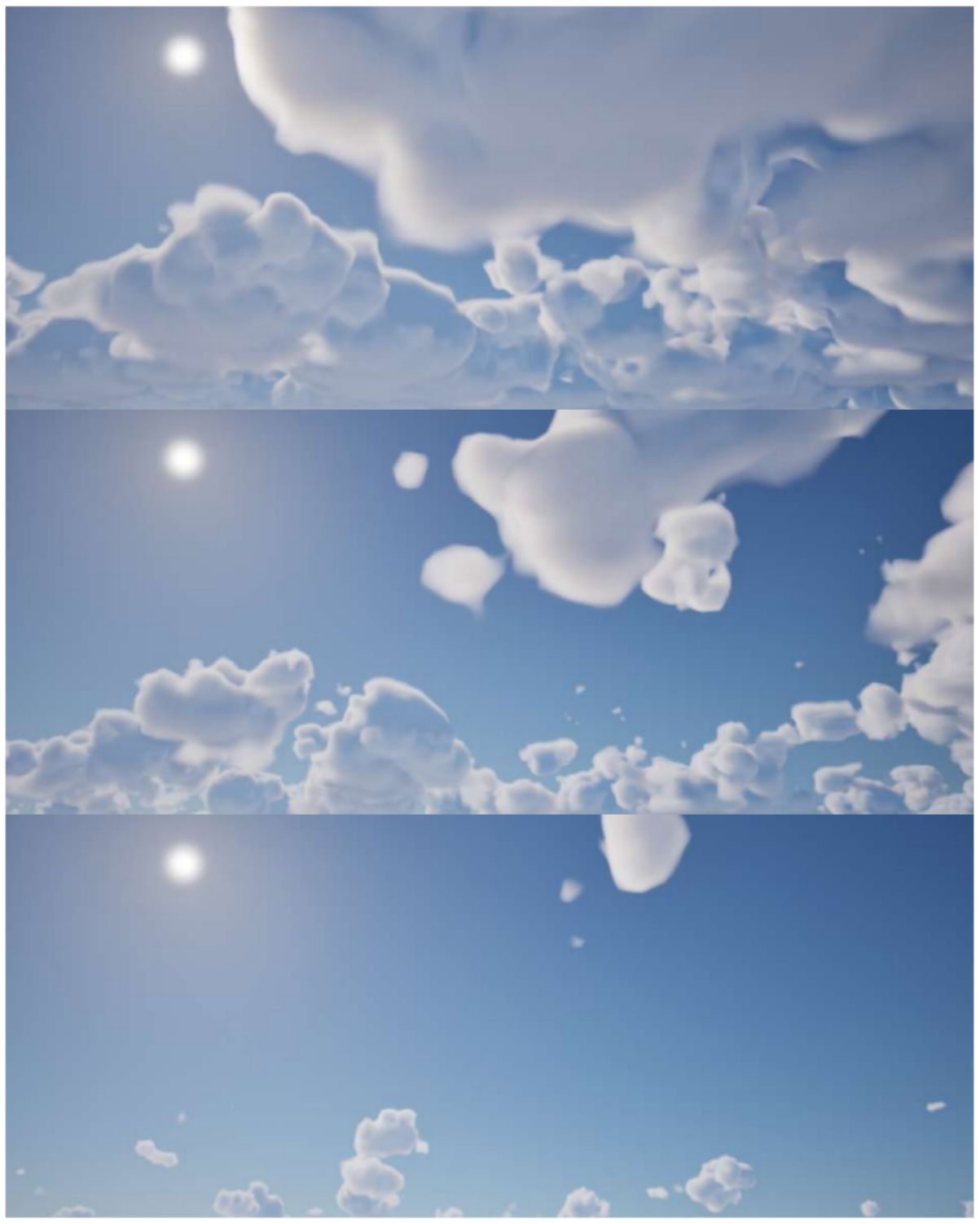

Fig. 3. $P_3 = 1.0$ is used in each render. $P_4 = 0.0$ for the top render, 0.4 for the middle render, and 1.2 for the bottom render

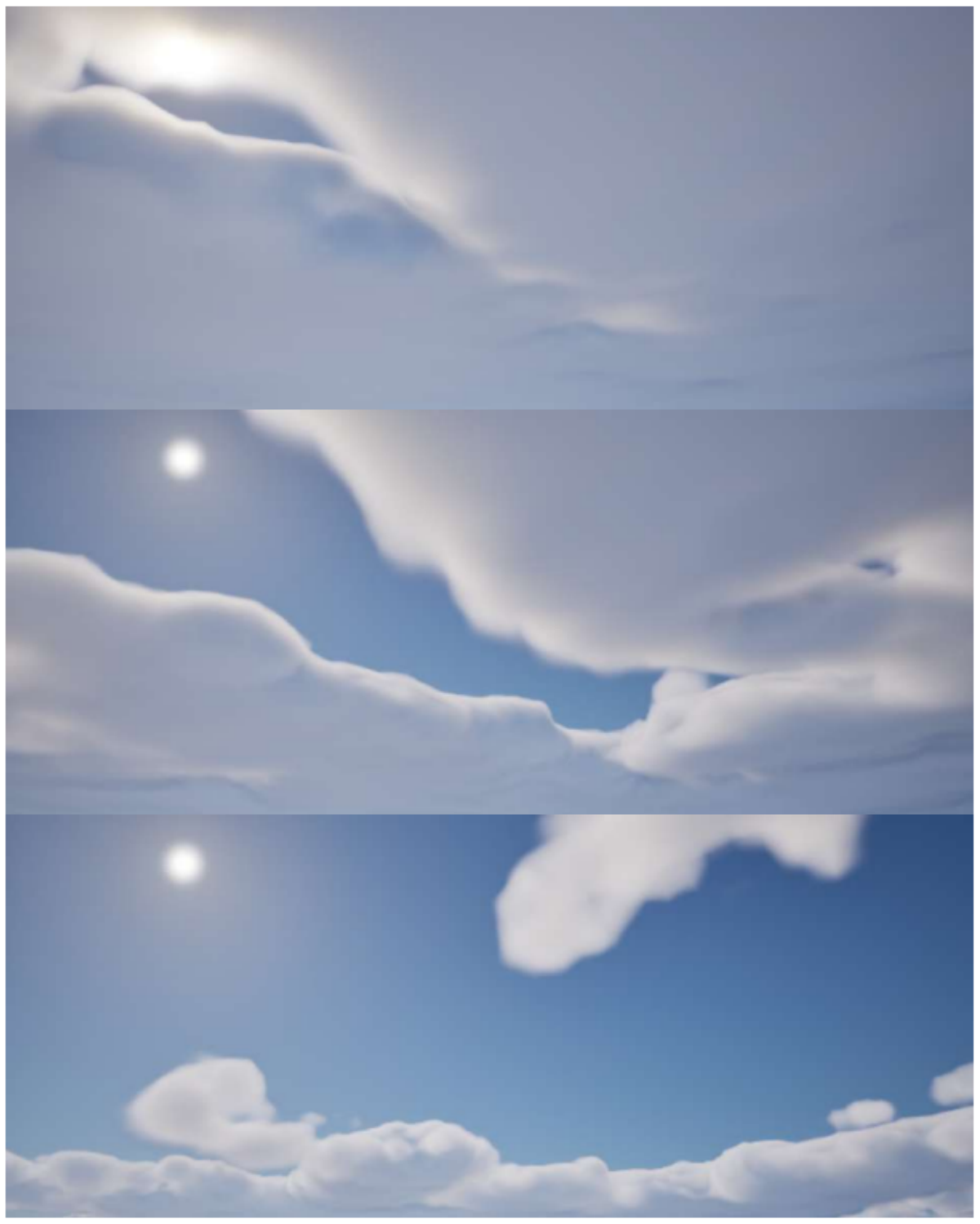

Fig. 4. $P_3 = 1.0$ is used in each render. $P_4 = 0.8$ for the top render, 1.0 for the center render, and 1.2 for the bottom render

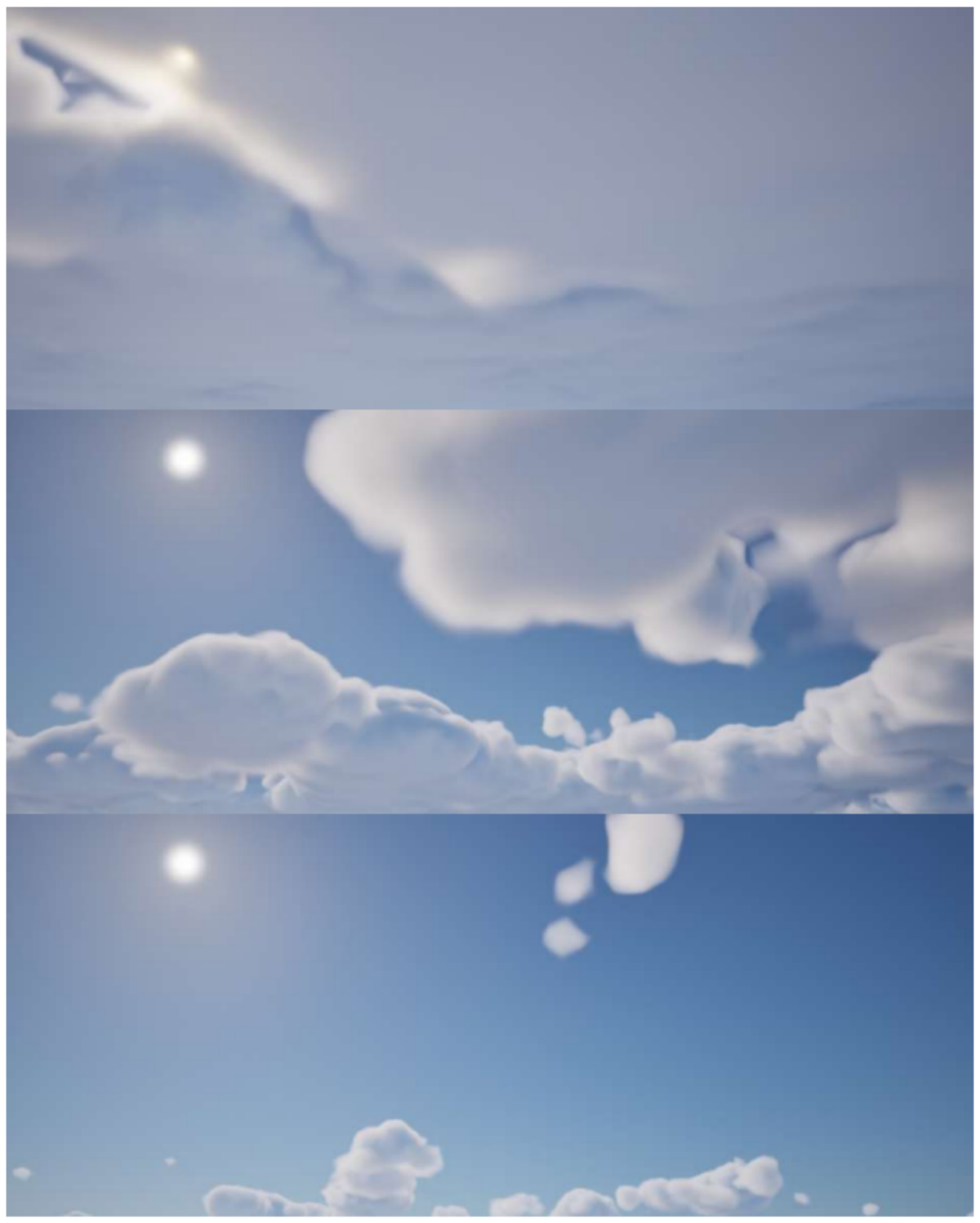

Fig. 5. The variables $P_3 = 1.0$, $C_{type}$[16] $= 0.024$, $C_{wispy}$[17] $= 0.248$, and $C_{billowy}$[18] $= 0.016$ are used in each render. $P_4 = 0.4$ for the top render, 0.85 for the middle render, and 1.2 for the bottom render

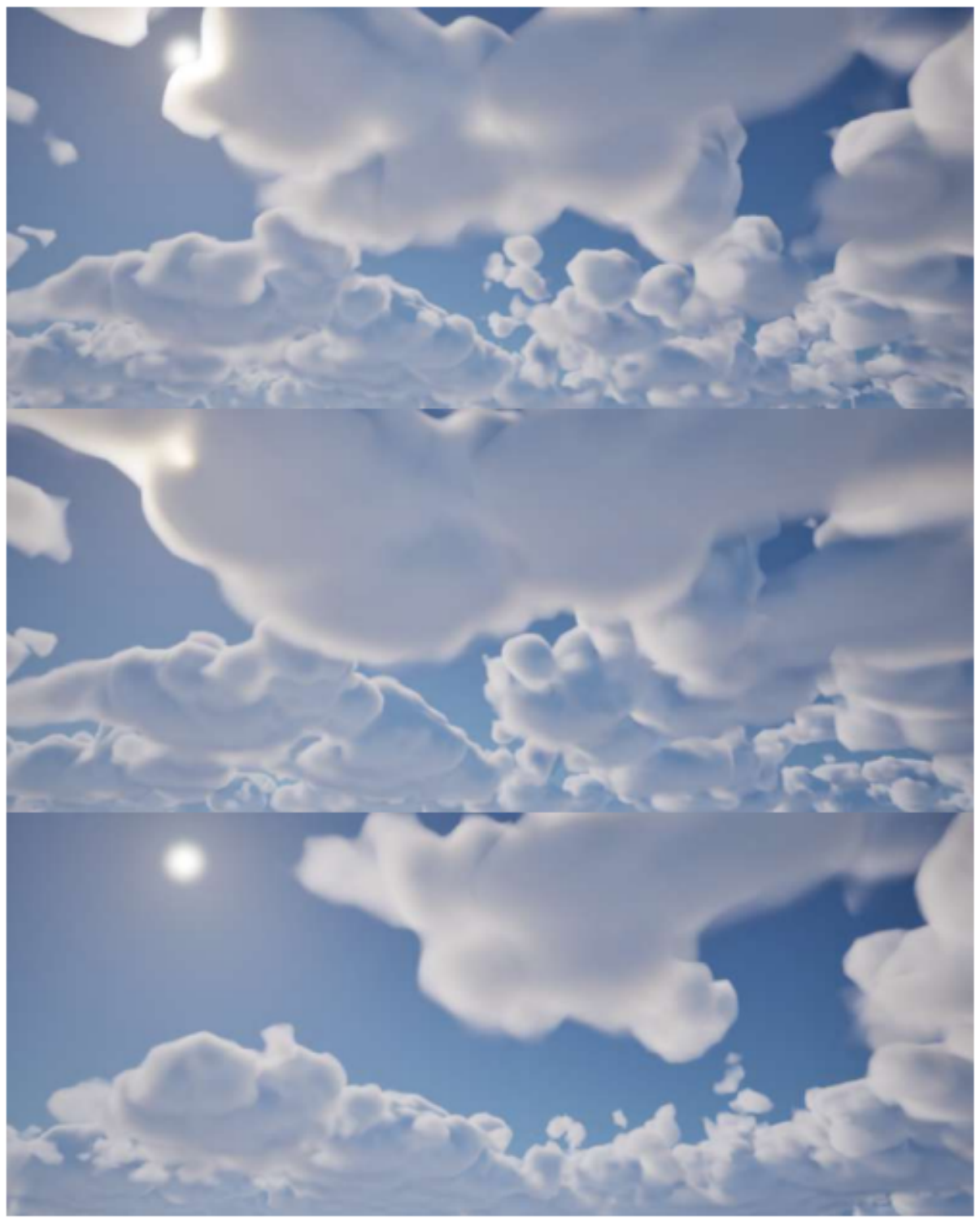

Fig. 6. The numbers $P_3 = 1.0$ and $P_4 = 0.85$ are used in each render. The renders at the top, middle, and bottom are $C_{type} = 0.5$, $C_{wispy} = 0.9$, and $C_{billowy} = 0.1$; $C_{type} = 0.2$, $C_{wispy} = 0.1$, and $C_{billowy} = 0.9$; and $C_{type} = 0.5$, $C_{billowy} = 0.9$, and $C_{billowy} = 0.1$

### *C. Erosion Details*

In the preceding section, we introduced a noise layer model that procedurally generates the base shapes of clouds. In this section, we explore our proposal for defining a second noise layer model, which introduces finer details to enhance the complexity of cloud structures. The concept behind this model is to create an additional noise layer and subtract its details from the base cloud shapes. To model the erosion[19] details of the cloud shapes, we utilize the 3D Curly-Alligator noise texture. This texture is distributed throughout the volume of the cloud layer and sampled. By determining the position of the current pixel sample in world space, the frequency of the erosion noise volume texture can be adjusted. This allows for greater control over the intricacies of cloud erosion and refinement. This method proves effective in intuitively adjusting the scale of the noise texture applied to the base cloud shapes, granting finer control over the erosion details of the clouds. Given that erosion details constitute a new noise layer, pseudo-motion is also simulated within this layer by defining an offset vector and adding it to the sampled position. However, the pseudo-motion for this layer is scaled by a user-defined parameter. This provides additional flexibility in controlling the dynamics of the erosion noise layer. This approach enables the erosion details to inherit the pseudo-motion of the base cloud shapes while allowing for the adjustment of its intensity, ensuring that this noise layer can move at a different pace. Since erosion details represent more subtle aspects of the cloud formations, it is appropriate for these details to represent less dense cloud structures and to move at a faster rate compared to the denser base cloud formations, as illustrated in Fig. 7. For realistic cloud modeling, it is essential that these erosion details remain subtle. Consequently, the tiling factor[20], must be significantly lower; otherwise, unrealistic cloud shapes will be rendered, as demonstrated in Fig. 8. This ensures that the erosion noise blends seamlessly with the base shapes, enhancing the overall cloud realism without introducing exaggerated features.

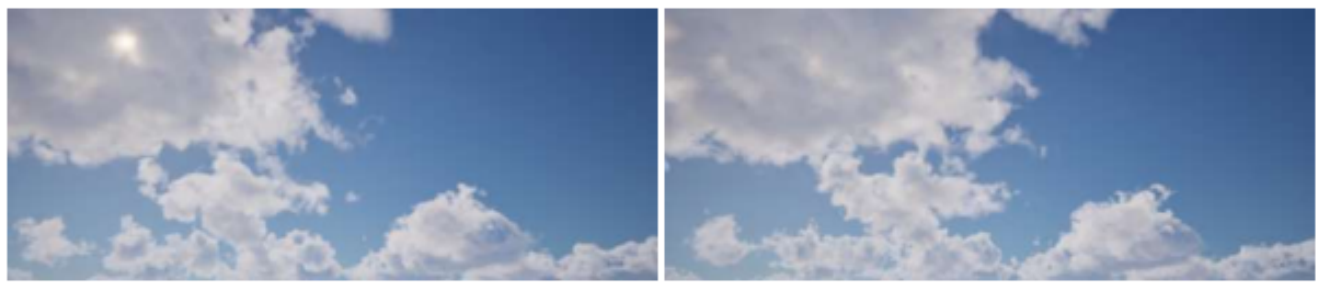

Fig. 7. Erosion details apply pseudo-motion to the cloud's basic shape

[16] Cloud Type Factor
[17] Wispiness Factor
[18] Density Factor

[19] By producing intricate and varied terrain features, erosion in computer graphics enhances realism by simulating the wearing away of surfaces.
[20] In 3D rendering, the tiling factor affects visual density and detail by determining how frequently a texture repeats across a surface.

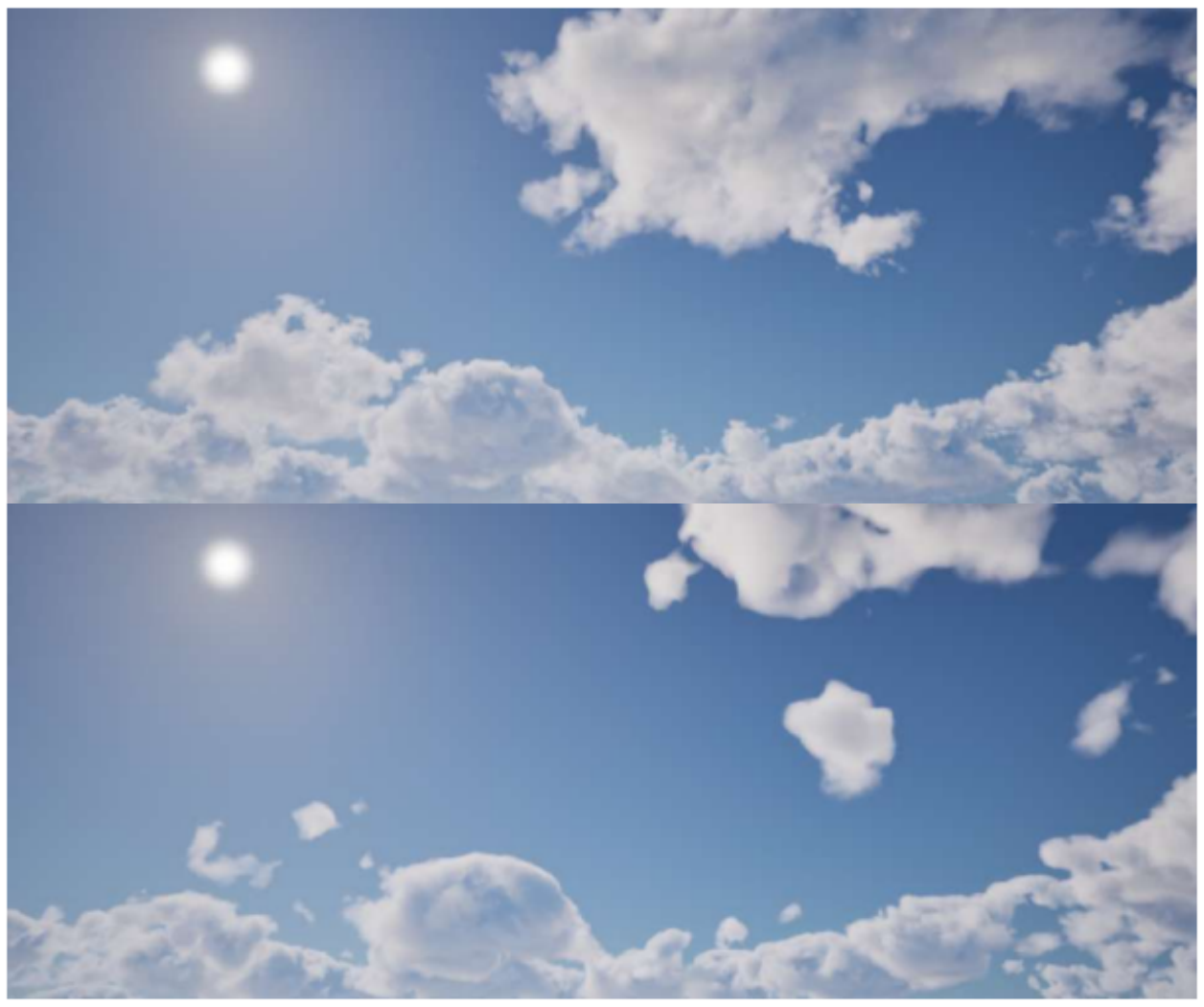

Fig. 8. In both representations, $b_{tiling}$ is equal to 30.0 km. The top render's $e_{tiling}$ is 3.8 km, whereas the bottom render's is 10.0 km

## V. Experimental Analysis

### A. Experimental Setup

In this study, several hyperparameters were systematically fine-tuned to enhance the accuracy and performance of our cloud rendering model. As shown in Table II, anisotropy (*g*), blending factor, particle size (*d*), and erosion frequency were adjusted based on empirical results and prior research. The anisotropy parameter, which controls the asymmetry of light scattering, was evaluated across a range from 0.5 to 0.9. A value of 0.85 was selected, as it yielded the most realistic cloud illumination while balancing rendering speed. The blending factor in the HG+D phase function, tested from 0.6 to 0.8, was fine-tuned to 0.7 to capture diverse lighting conditions without overexposing or dulling the cloud edges. For noise texture, the erosion frequency was set to 1.5 after several tests demonstrated its impact on cloud depth and softness, enabling better visual distinction among cloud types. Lastly, particle size (*d*) was fine-tuned at 4.5 to mimic real-world particle distributions and enhance light interaction, yielding more life-like cloud formations. These optimizations, detailed in Table II, align the model's psarameters with observed natural cloud behavior, improving visual authenticity without compromising computational efficiency. Each hyperparameter was carefully balanced to allow flexibility in generating multiple cloud types with various atmospheric properties.

TABLE II
HYPERPARAMETERS

| Hyperparameter | Range Tested | Final Value |
|---|---|---|
| Anisotropy (g) | 0.5 - 0.9 | 0.85 |
| Blending Factor | 0.6 - 0.8 | 0.7 |
| Particle Size (d) | 3.0 - 5.0 | 4.5 |
| Erosion Frequency | 1.0 - 2.0 | 1.5 |
| Base Noise Frequency | 1.0 - 1.5 | 1.2 |

### B. Datasets

The dataset for training and evaluating our cloud rendering model consists of procedurally generated volumetric cloud scenes, utilizing advanced particle-based rendering and noise textures to simulate diverse atmospheric conditions. Table III highlights dataset specifications, including 500 scenes of various cloud types such as cirrus, cumulus, and stratocumulus, each tested with multiple phase functions and lighting angles to ensure robustness across conditions. Each cloud scene is captured at a 512×512-pixel resolution to maintain a high level of detail while optimizing for computational efficiency. Scenes are further categorized by phase functions, including TTHG and HG+D, which are applied to simulate varying scattering properties across cloud formations. A 70%-30% train-test split was chosen to support robust model validation and reduce overfitting, particularly for scenes with complex lighting and structural variations. Each sample includes metadata for light angle, cloud density, and altitude, ensuring that the model can generalize across a wide range of environmental settings.

TABLE III
DATASET ANALYSIS

| Dataset Parameter | Value |
|---|---|
| Total Samples | 500 |
| Resolution | 512×512 pixels |
| Cloud Types | Cirrus, Cumulus, Stratocumulus |
| Phase Functions | TTHG, HG+D |
| Train-Test Split | 70%-30% |
| Lighting Angles | Varied |

### C. Main Results

In the preceding subsections, all rendered images employed consistent settings for ray casting views and maximum shadow samples: a total of $760 \times 4.0 = 3{,}072$ view samples and $80 \times 4.0 = 320$ shadow samples, accompanied by a mean tracing transmittance threshold of 0.005. The rendered images presented in these sections served benchmarking purposes. Our proposed solution was implemented as a shader designed to execute on a GPU. For benchmarking, we tested our solution across various GPUs, maintaining a resolution of 1080p (Full HD), and the results are documented in Table IV and Fig. 9. The displayed data denote the average performance metrics (measured in milliseconds) of the VCC, as analyzed using the GPU profiler integrated within UE. Notably, the GPU RTX 2070 Super 8GB was a laptop version, while the remaining GPUs were all desktop variants. The selection of diverse GPUs for testing provided insights into performance across different hardware generations. Analysis of the data reveals that rendering scenes with the sun positioned on the horizon, closely aligned with the cloud layer, results in a significant increase in performance costs. This increase occurs because the sampling within the shadow rays must cover the entire length of the ray, often necessitating the maximum sample count set at 320. The chosen number of samples enabled the generation of high-quality, realistic cloud shapes without noticeable artifacts when viewed from the ground. By reducing the sample count to half, with both view sample count scale = 2.0 and shadow view sample count scale = 2.0, performance can improve significantly, as documented in Table and Fig. 10. However, this reduction may introduce minor artifacts in certain scenes. The complexity of the shader is primarily influenced by the number of texture reads and the proportion that the sky occupies within the renders. Thus, the careful selection of the user parameters is essential, as these directly impact performance. Lower values lead to increased texture reads, as the texture will be tiled multiple times within the same render. Additionally, less dense cloud structures tend to perform worse than denser formations in terms of rendering performance. This is due to the ray-marching algorithm reaching a maximum transmittance computation sooner in denser structures, resulting in fewer computational steps required for accurate rendering. The height of the cloud layer also warrants careful consideration, as it directly affects ray lengths, thereby influencing performance.

TABLE IV
COMPARATIVE PERFORMANCE OF TTHG AND HG+D PHASE FUNCTIONS FOR CLOUD RENDERING IN UE

| **Parameter** | **TTHG Phase Function** | **HG+D Phase Function** | **Comments** |
|---|---|---|---|
| Anisotropy ($g$) | 0.75 | 0.85 | Higher values produce brighter effects near light sources |
| Blending Factor | 0.65 | 0.78 | Controls blending between HG and Dirac phase functions |
| Particle Size ($d$) | 3.5 | 4.7 | Larger particles result in increased light scattering |
| In-Scattering Intensity | 0.85 | 0.65 | HG+D aligns with Mie scattering, yielding moderate in-scattering |
| Out-Scattering Effect | 0.75 | 0.85 | Enhanced out-scattering in HG+D provides realistic cloud edges |
| Rendering Time (ms) | 30 | 45 | HG+D requires more computation time |
| Visual Accuracy | Moderate | High | HG+D closely aligns with physical Mie scattering |
| Flexibility | High | Moderate | TTHG allows more customizable artistic control |

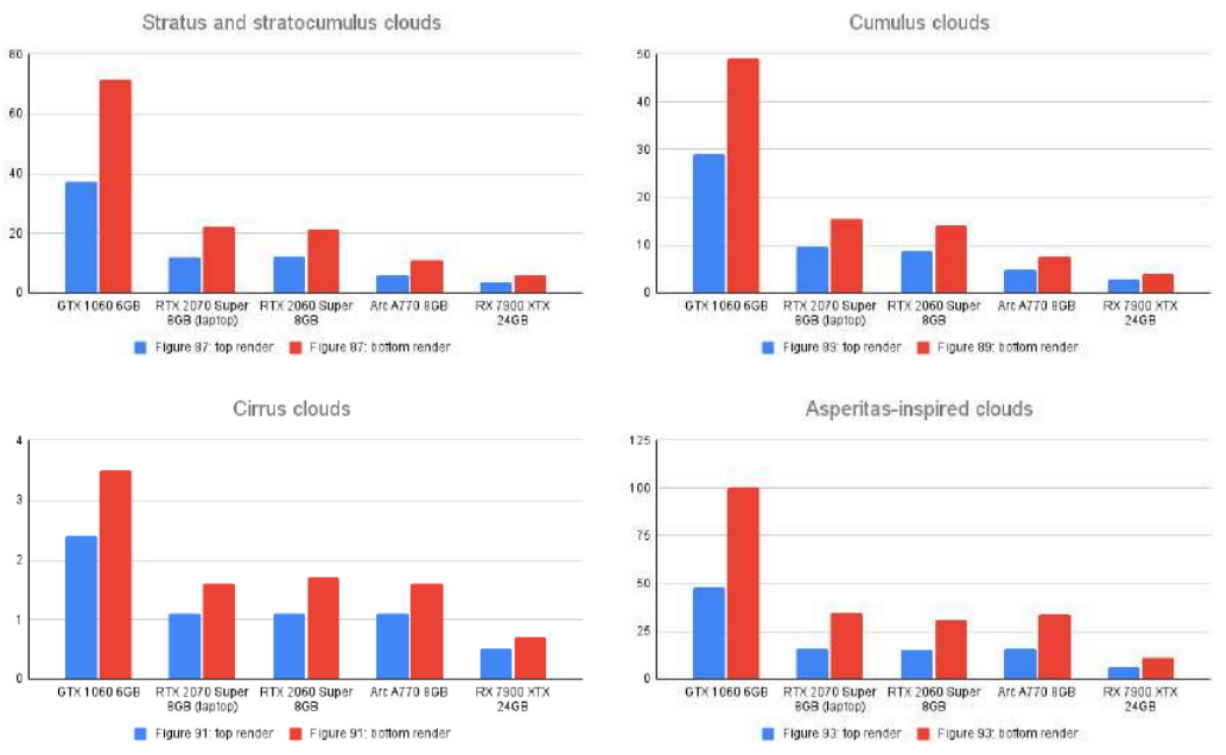

Fig. 9. Graphs showing the benchmark frames (in milliseconds) on the GPUs under test

TABLE V
PROCEDURAL NOISE TEXTURE PARAMETERS AND CLOUD SHAPE DETAILS FOR DIFFERENT APPROACHES

| **Parameter** | **Method 1 (Noise Composite)** | **Method 2 (Alternative Shaping)** | **Method 3 (Sequential Interpolation)** | **Comments** |
|---|---|---|---|---|
| Base Noise Frequency | 1.25 | 1.35 | 1.15 | Controls cloud density distribution |
| Erosion Frequency | 1.4 | 1.6 | 1.7 | Adds structural detail and depth |
| Coverage Control (P4) | 0.55 | 0.8 | 1.1 | Modifies cloud coverage over the scene |
| $C_{type}$ (Cloud Type Factor) | 0.4 | 0.25 | 0.5 | Adjusts rendering for visual diversity |
| $C_{wispy}$ (Wispiness Factor) | 0.85 | 0.3 | 0.9 | Higher values generate wispy textures |
| $C_{billowy}$ (Density Factor) | 0.2 | 0.9 | 0.3 | Higher values indicate denser cloud forms |
| Memory Usage (MB) | 185 | 215 | 245 | Sequential interpolation requires more memory |
| Render Time (ms) | 26 | 35 | 43 | Method 3 requires more processing time |
| Realism | Medium | High | Very High | Method 3 provides smoother transitions |

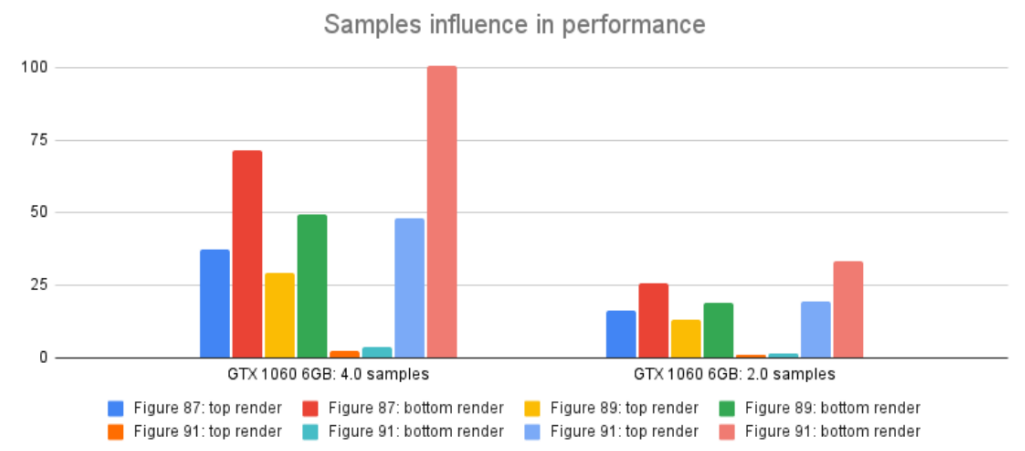

Fig. 10. Graph that compares benchmarks for the GPU GTX 1060 6GB and illustrates the impact of sample size selection

## VI. Conclusion and Future Works

In this study, we have examined the challenges and techniques associated with real-time volumetric cloud rendering. Our primary objectives were to investigate the complexities related to cloud modeling and rendering, alongside the state-of-the-art methods aimed at addressing these challenges. We analyzed approaches centered on ray marching algorithms, volumetric textures, and procedural methods. The current state-of-the-art cloud modeling procedures are predominantly artistically driven, relying on secondary 2D weather textures to define cloud placement within the cloud layer. Such textures not only necessitate additional texture reads for the algorithms but also demand artistic skills for their manual creation. We proposed a novel solution for cloud modeling implemented in UE, utilizing procedural methods and volumetric textures to define two distinct noise layers, thereby eliminating the requirement for a secondary 2D weather texture for cloud generation. Three different methods were developed to define cloud placement, complemented by several user parameters that enable control over the various components of our model. This allows for the configuration of a dynamic sky using analytical methods. Our solution successfully renders a diverse array of realistic cloud shapes, validating the effectiveness of our approach. The implemented solution achieves compelling visual results, facilitating the generation of fully procedural skies in real time. Cloud animation was accomplished through the application of simple pseudo-motion. By utilizing two distinct noise layers—one defining the base shape of the clouds and the other detailing its nuances—it was possible to simulate motion by animating the details layer at a faster rate than the base layer. This method yields satisfactory results for basic cloud animations. The utilization of UE's framework was paramount, providing state-of-the-art physically based techniques for cloud lighting and rendering. This choice has proven to be a robust and effective solution, serving as a compelling case study.

We also proposed extending the existing solution within UE to incorporate phase function sampling. Various studies have suggested differing values for the effective cloud radius, making this solution straightforward to implement for physically based simulations, as it relies on a single parameter: droplet diameter. In terms of performance, real-time simulations must be finely tuned to meet specific requirements based on the simulation budget. Our solution implemented in UE has demonstrated commendable performance across various generations of GPU hardware. We focused on generating high-quality and realistic cloud shapes, achieving notably good performance, especially on the latest hardware configurations. The primary concern for performance resides within the ray-marching algorithm, which necessitates careful fine-tuning. Testing was conducted under conditions devoid of obstacles in the sky to properly evaluate performance in extreme scenarios where the sky encompasses the entire screen. Our solution was rigorously tested under these extreme conditions and has exhibited considerable potential across a broad range of performance budgets, highlighting its adaptability to different consumer hardware configurations. Through meticulous optimization of the ray-marching algorithm, we can produce high-quality and realistic cloud shapes from ground-level perspectives on any tested hardware configuration. Thus, we believe our solution has proven successful for real-time simulations.

## VII. Declarations

*A.* ***Funding:*** No funds, grants, or other support was received.

*B.* ***Conflict of Interest:*** The authors declare that they have no known competing for financial interests or personal relationships that could have appeared to influence the work reported in this paper.

*C.* ***Data Availability:*** Data will be made on reasonable request.

***D. Code Availability:*** Code will be made on reasonable request.

## References

[1] "Advances in Real-Time Rendering- SIGGRAPH 2015." Accessed: Oct. 27, 2024. [Online]. Available: https://advances.realtimerendering.com/s2015/index.html

[2] W. Engel, *GPU pro: Advanced rendering techniques*. A K Peters/CRC Press, an imprint of Taylor and Francis, 2010. doi: 10.1201/b10648.

[3] C. R. Nave and J. Sheridan, "The microwave and infrared spectra and structure of hydrothiophosphoryl difluoride," *J. Mol. Struct.*, vol. 15, no. 3, pp. 391–398, Mar. 1973, doi: 10.1016/0022-2860(73)80008-0.

[4] G. Mie, "Beiträge zur Optik trüber Medien, speziell kolloidaler Metallösungen," *Ann. Phys.*, vol. 330, no. 3, pp. 377–445, Jan. 1908, doi: 10.1002/andp.19083300302.

[5] Y. A. Eremin, "Scattering: Scattering Theory," in *Encyclopedia of Modern Optics, Five-Volume Set*, Elsevier, 2004, pp. 326–330. doi: 10.1016/B0-12-369395-0/00682-5.

[6] A. Zucconi, "An Introduction to Neural Networks and Autoencoders." Accessed: Oct. 27, 2024. [Online]. Available: https://www.alanzucconi.com/2020/07/30/atmospheric-scattering-8/

[7] L. G. Henyey and J. L. Greenstein, "Diffuse radiation in the galaxy," 1941. doi: 10.1086/144246.

[8] T. Nishita, T. Sirai, K. Tadamura, and E. Nakamae, "Display of the earth taking into account atmospheric scattering," in *Proceedings of the 20th Annual Conference on Computer Graphics and Interactive Techniques, SIGGRAPH 1993*, Association for Computing Machinery, Sep. 1993, pp. 175–182. doi: 10.1145/166117.166140.

[9] A. Zucconi, "The Mathematics of Rayleigh Scattering.," 2017. Accessed: Oct. 27, 2024. [Online]. Available: https://www.alanzucconi.com/2017/10/10/atmospheric-scattering-3/

[10] T. Nishita, Y. Dobashi, and E. Nakamae, "Display of clouds taking into account multiple anisotropic scattering and sky light," in *Proceedings of the 23rd Annual Conference on Computer Graphics and Interactive Techniques, SIGGRAPH 1996*, Association for Computing Machinery, Inc, Aug. 1996, pp. 379–386. doi: 10.1145/237170.237277.

[11] A. J. Preetham, P. Shirley, and B. Smits, "A practical analytic model for daylight," in *Proceedings of the 26th Annual Conference on Computer Graphics and Interactive Techniques, SIGGRAPH 1999*, Association for Computing Machinery, Inc, Jul. 1999, pp. 91–100. doi: 10.1145/311535.311545.

[12] R. Perez, R. Seals, and J. Michalsky, "All-weather model for sky luminance distribution-Preliminary configuration and validation," *Sol. Energy*, vol. 50, no. 3, pp. 235–245, Mar. 1993, doi: 10.1016/0038-092X(93)90017-I.

[13] "Talk at 2003 Game Developers Conference." Accessed: Oct. 27, 2024. [Online]. Available: https://renderwonk.com/publications/gdc-2002/

[14] R. S. Nielsen, "Real Time Rendering of Atmospheric Scattering Effects for Flight Simulators," no. Imm, p. 134, 2003, Accessed: Oct. 27, 2024. [Online]. Available: https://www2.compute.dtu.dk/pubdb/pubs/2554-full.html

[15] J. Haber, M. Magnor, and H. P. Seidel, "Physically-based simulation of twilight phenomena," *ACM Trans. Graph.*, vol. 24, no. 4, pp. 1353–1373, Oct. 2005, doi: 10.1145/1095878.1095884.

[16] C. Sigg and M. Hadwiger, "Fast third-order texture filtering," *GPU Gems 2 Program. Tech. High-Performance Graph. Gen. Comput.*, pp. 313–329, 2005, Accessed: Oct. 27, 2024. [Online]. Available: https://developer.nvidia.com/gpugems/gpugems2/part-ii-shading-lighting-and-shadows/chapter-16-accurate-atmospheric-scattering

[17] C. Wenzel, "Real-time atmospheric effects in games," in *SIGGRAPH 2006 - ACM SIGGRAPH 2006 Courses*, 2006, pp. 113–128. doi: 10.1145/1185657.1185831.

[18] T. Schafhitzel, M. Falk, and T. Ertl, "Real-time rendering of planets with atmospheres," in *15th International Conference in Central Europe on Computer Graphics, Visualization and Computer Vision 2007, WSCG'2007 - In Co-operation with EUROGRAPHICS, Full Papers ProceedingsWSCG Proceedings*, 2007, pp. 91–98.

[19] E. Bruneton and F. Neyret, "Precomputed atmospheric scattering," *Comput. Graph. Forum*, vol. 27, no. 4, pp. 1079–1086, 2008, doi: 10.1111/j.1467-8659.2008.01245.x.

[20] O. Elek, "Rendering Parametrizable Planetary Atmospheres with Multiple Scattering in Real-Time," *Cescg*, 2009.

[21] E. Yusov, "Outdoor Light Scattering Sample Update," 2013, Accessed: Oct. 27, 2024. [Online]. Available: http://software.intel.com/en-us/blogs/2013/06/26/outdoor-light-

scattering-sample

[22] L. Hosek and A. Wilkie, “An analytic model for full spectral sky-dome radiance,” *ACM Trans. Graph.*, vol. 31, no. 4, Jul. 2012, doi: 10.1145/2185520.2185591.

[23] “SIGGRAPH 2013 Course: Physically Based Shading in Theory and Practice.” Accessed: Oct. 27, 2024. [Online]. Available: https://blog.selfshadow.com/publications/s2016-shading-course/

[24] J. Jendersie and E. D’Eon, “An Approximate Mie Scattering Function for Fog and Cloud Rendering,” in *Proceedings - SIGGRAPH 2023 Talks*, Association for Computing Machinery, Inc, Aug. 2023. doi: 10.1145/3587421.3595409.

[25] B. T. Draine, “Scattering by Interstellar Dust Grains. I. Optical and Ultraviolet,” *Astrophys. J.*, vol. 598, no. 2, pp. 1017–1025, Dec. 2003, doi: 10.1086/379118.

[26] “SIGGRAPH 2023 Advances in Real-Time Rendering in Games course.” Accessed: Oct. 27, 2024. [Online]. Available: https://www.advances.realtimerendering.com/s2023/

[27] “Authoring Pseudo Volume Textures.” Accessed: Oct. 27, 2024. [Online]. Available: https://shaderbits.com/blog/authoring-pseudo-volume-textures